\documentstyle[aps,preprint,tighten,epsfig,amssymb]{revtex}

\begin{document}


\draft


\title{
An isotopic effect in  $\bbox{\phi}$ photoproduction
at a few GeV 
}


\author{
A.~Titov\,$^{a,b}$%
\footnote{E-mail address : {\tt atitov@thsun1.jinr.ru}},
T.-S.H.~Lee\,$^c$%
\footnote{E-mail address : {\tt lee@anph09.phy.anl.gov}}
and
H.~Toki\,$^a$%
\footnote{E-mail address : {\tt toki@rcnp.osaka-u.ac.jp}},}


\address{
$^a$ Research Center for Nuclear Physics, Osaka University, Osaka 567-0047,
Japan\\
$^b$ Bogoliubov Laboratory of Theoretical Physics, JINR, Dubna 141980,
Russia\\
$^c$ Physics Division, Argonne National Laboratory, Argonne,IL 60439,  USA
}

\maketitle

\begin{abstract}
A distinct isotopic effect in $\phi$ photoproduction 
at $\sqrt{s}\sim 2-5$ GeV region is identified by examining 
the production amplitudes due to Pomeron-exchange and meson-exchange
mechanisms. 
This effect is mainly caused by the
$\pi$-$\eta$ interference constrained by SU(3) symmetry and
the isotopic structure  of the $\gamma NN$ coupling in the direct
$\phi$-radiation amplitude.
It can be tested experimentally by measuring differences in
the polarization observables between the $\gamma p\to \phi p$
and  $\gamma n\to \phi n$ reactions.
\end{abstract}

\pacs{PACS number(s): 13.88.+e, 24.70.+s, 25.20.Lj, 13.60.Le}


${\bbox{\rm Introduction.}}$
Traditionally, the $\gamma p\to \phi p$ reaction is considered
as a tool to study the Pomeron exchange dynamics,
implying that (i) the vector meson dominance
model (VDM)~\cite{DL84-86,PL96} is valid,
and (ii) other hadronic mechanisms are  
suppressed by OZI rule. The Pomeron-exchange mechanism is depicted
in Fig.1a.
Together with an examination of the conventional Pomeron trajectory
with an intercept
$\alpha(0)\sim 1.08$~\cite{PL96,DL87a,LN87,DL88-92,LM95},
it has been suggested that  this reaction can be used
to probe more exotic processes associated with additional trajectories
including the glueball trajectory~\cite{NTL}.

In order to learn about the
Pomeron structure and possible manifestation of more exotic dynamics,
such as the hidden strangeness in proton
(cf. \cite{Ellis,TOYM98} for references),
one should first fix the ``non-strange'' background determined by
the conventional mechanisms. Namely, we need to evaluate 
as accurately as possible the leading amplitudes due to 
the pseudo-scalar ($\pi,\,\eta$)-exchange (PSE) and the
direct $\phi$-radiation from the nucleon legs.
They are illustrated in Fig.1b and Fig.1c respectively.
The $\eta$-exchange and direct $\phi$-radiation mechanisms are of
current interest.
In particular, the
coupling constant $g_{\eta NN}$ is not well established
and it is subject to further investigations. Similarly,
our understanding of $\phi NN$ coupling is still limited.
Its OZI-violation value has been evaluated, for example 
in Ref.\cite{MMSV}, by considering 
the finite $\phi \pi\rho$ couplings and the
interaction of the $\phi$-meson with the  kaon cloud of the nucleon.

A step towards an understanding of the structure of the ``non-strange''
background in $\phi$ photoproduction is taken in a recent
paper~\cite{Will98}.
The focus of Ref.~\cite{Will98}
is to explore the possibility of determining the
$\phi NN$ and $\eta NN$ coupling constants using the
$\gamma p\to \phi p$ reaction. Except the  $\pi NN$
coupling constant, all parameters associated with Pomeron-exchange,
$\eta$ exchange, and direct $\phi$-radiation amplitudes 
are adjusted to fit the data.
In this paper we take a different approach. We calculate
the $(\pi,\eta)-$exchange, and the direct $\phi-$radiation
amplitudes 
using the parameters {\it predetermined} in previous works
constrained by SU(3) symmetry.
This will allow us to pin down
the parameters associated with an additional Pomeron trajectory
that will be introduced later.

In Ref.~\cite{TLTS} we give a detailed analysis of $\phi$ photoproduction
amplitudes in a few GeV region and predict all spin density matrix
elements that are needed to evaluate various observables
associated with the $K^+K^-$ angular distributions from the
decay of the produced $\phi$ meson.
In the present Letter we emphasis an
interesting isotopic effect caused by the  $\pi$-$\eta$ interference(Fig.1b) 
and the difference between the photon-proton and photon-neutron
vertices in the direct $\phi$-radition mechanism(Fig.1c).
Experimental tests of our predictions will help pin down the
"non-strange" background of $\phi$ photoproduction.

\noindent
We define the kinematical variables for 
the $\gamma + N \to \phi + N$ reaction using standard
notations.
The four-momenta of the incoming photon, outgoing $\phi$, initial
nucleon, and final nucleon are denoted as $k$, $q$, $p$, and $p'$,
respectively. The  usual Mandelstam variables are then defined as
$t = (p - p')^2=(q-k)^2$ and $s\equiv W^2 = (p+k)^2$. The masses of
the nucleon, pion, and $\phi$ meson are $M_N$,
$M_\pi$, and $M_\phi$ respectively.
We use the conventions of Bjorken\&Drell to define
the $\gamma$ matrices and Dirac spinors.

\noindent
$\bbox{\rm Pomeron\, exchange\, amplitudes.}$
A microscopic model for vector-meson photo- and electro-production
at high energies based on the Pomeron-photon analogy has been proposed
by Donnachie and Landshoff \cite{DL87a} (DL-model).
It is assumed that
the incoming photon first converts into a quark-antiquark pair, 
which then exchanges a Pomeron
with the nucleon before recombining into an outgoing
$\phi$ meson, as depicted in Fig.~2.
It can be shown that the data at high energies can be described by
a Pomeron trajectory with an intercept $\alpha(0) \sim 1.08$.
This trajectory will be called the $P_1$ trajectory in this paper.
Within Regge theory, it is easy to see that the P$_1$ trajectory
can only be identified with objects with spin $J>0$. 
It was suggested ~\cite{NTL}
that a different Pomeron trajectory,
as inspired by the glueball
($J^\pi=0^+$, $M^2_{\rm gb}\simeq 3$ GeV) 
predicted by
Lattice QCD calculations and a QCD-based Dual Landau-Ginsburg
model~\cite{LGM}, could dominate the cross section at low energies
by having a negative intercept $\alpha(0)$. This Pomeron trajectory,
called P$_2$, will be also considered in this work.

The DL-model  was examined in Ref.~\cite{PL96} by using an Euclidean
QCD model of
mesons to evaluate exactly the quark-loop integration in Fig.2. It was shown
that the interpretation of the Pomeron as gluonic object is consistent
with the data up to very large $Q^2$ and $W$.
For the $Q^2=0$ and $W<5$ GeV region considered in this work,
the exact loop integration, as done in Ref.~\cite{PL96},
can be approximated by the factorized form employed by
Donnachie and Landshoff. We therefore follow Donnachie and Landshoff
and use this simplified version of the model 
to define the Pomeron-exchange amplitudes.
We will not consider two-gluon-exchange models of Pomeron as
proposed in Refs.~\cite{LN87,DL88-92,Cud90}.

Within the DL-model, the invariant amplitude can be written as
\begin{equation}
T_{fi}^{\rm P_{n}} = M_0^{\rm P_{n}}
\varepsilon^*_{\phi\mu}\, {{\cal M}^{\rm P_{n}}}^{\mu\nu}\,
\varepsilon_{\gamma\nu},\quad\quad
{{\cal M}^{P_n \mu\nu}} = {\cal F}^{P_n}_\alpha \, \Gamma^{P_n \alpha,\mu\nu},
\label{T-VDM}
\end{equation}
where $\varepsilon_{\gamma\nu}$ and $\varepsilon_{\phi\mu}$
are respectively the polarization vectors of the photon and $\phi$ meson,
{${\cal F}^{P_n}_\alpha$ describes the Pomeron-nucleon vertex, and
$\Gamma^{P_n  \alpha,\mu\nu}$ is associated with the Pomeron--vector-meson
coupling which is related to the $\gamma \to q \bar q$ vertex $\Gamma_\nu$
and the $q\bar q \to \phi$ vertex $V_\mu$, as shown in Fig.~2.
We follow the DL-model and assume that the P$_1$-nucleon coupling is
photon-like. For the P$_2$-nucleon vertex, we assume a scalar coupling.
Accordingly, we have
\begin{equation}
{\cal F}^{P_n}_\alpha = \bar u (p') \,g^{P_n}_\alpha u(p),\qquad
g^{\rm P_1}_\alpha =\gamma_\alpha,\,\, \,\,
g^{\rm P_2}_\alpha =1,
\label{Fal}
\end{equation}
where $u(p)$ is the nucleon spinor with momentum $p$.
By using the factorization approximation of Donnachie and Landshoff,
an explicit evulation~\cite{TOYM98} of the Dirac algebra associated
with the quark-loop shown in Fig.2 leads to the following form
of the Pomeron-vector meson coupling in Eq.(1)
\begin{eqnarray}
\Gamma^{{\rm P_1}\alpha,\mu\nu} =
 g^{\alpha\nu}\, k^\mu -
k^\alpha \,g^{\mu\nu},\qquad\qquad
\Gamma^{{\rm P_2}\mu\nu}=
({ k^\mu q^\nu} - k\cdot q\, g^{\mu\nu})/M_\phi ,
\label{Gam2}
\end{eqnarray}
where the transversality conditions
${{\cal M}^{\rm P_{1,2}}}^{\mu\nu}\cdot q_\mu =
{{\cal M}^{\rm P_{1,2}}}^{\mu\nu}\cdot k_\nu=0$ are fullfilled.
The strength factors $M_0^{\rm P_{1,2}}$ in Eq.(1)
take the conventional form of Regge parameterization
\begin{eqnarray}
&&M_0^{\rm P_{i}}=C_i\,F(t)\, F_i(s,t)\,
e^{-i\frac{\pi}{2}\alpha_i(t)}\,
\left( \frac{s-s_i}{s_0} \right)^{\alpha_i(t)},
\label{M0P2}
\end{eqnarray}
where the trajectories are taken to be
\begin{eqnarray}
&&\alpha_1(t)=\alpha_1(0) + \alpha_1'\,t = 1.08 + 0.25\,t,\qquad
\alpha_2(t)= -0.75 + 0.25\,t,
\quad s_0={\alpha'}^{-1}.
\end{eqnarray}
As discussed above, $\alpha_1(t)$ is chosen to fit the data at large $s$ and
$\alpha_2(t)$ is inspired by the ($J^\pi =0^+$,$ M_{gb}^2 \sim 3$ GeV$^2$)
glueball, predicted by Lattice QCD calculations and a QCD-based 
Dual Landau-Ginsbug model\cite{LGM}. 
We follow Ref.\cite{LM95} to define the 
overall form factor in Eq.(4) as $F(t)=F_\phi(t)\cdot F_N(t)$~\cite{LM95}
with
\begin{eqnarray}
&&F_N(t)=\left({M_N^2-2.8t}\right)/\left({(4M_N^2-t)(1-(t/0.7))^2}\right),
\quad
F_\phi(t)=\exp[B(t-t_{\rm max})].
\label{POMFF}
\end{eqnarray}
The correcting functions $F_{1,2}(s,t)$ for Eq.(4) are 
\begin{equation}
F_{i}^{-2}=
\Gamma_{\mu\nu}^{P_i\alpha}\, \Gamma_{\mu'\nu'}^{P_i\alpha'}
 \, \mbox{Tr}\,
\left\{ g^{P_i}_{\alpha} (\not\hskip-0.7mm\!{p} + M_N ) g^{P_i}_{\alpha'}
(\not\hskip-0.7mm\!{p'}  + M_N )   \right\}\,
(g^{\mu\mu'}-q^\mu q^{\mu'}/M_\phi^2)\,g^{\nu\nu'}/4M_N^2 .
\end{equation}
The strength factor $C_1$ in Eq.(4) is chosen to reproduce
$d\sigma/dt |_{\theta=0}$
at large $s$, where the cross section is entirely due to P$_1$
trajectory.
The strength factor $C_2$ is chosen
to reproduce  $d\sigma/dt|_{\theta=0}$
at low energies where all mechanisms shown in Fig.1 are important.
The slope  $B$ is chosen to reproduce the available data of
angular distributions, $d\sigma/dt$.
Note that the phase of the Pomeron exchange amplitudes is fixed and
controlled by the exponential factors in~(\ref{M0P2}).
At $t=0$, ${\rm P_1}$ amplitude is almost purely imaginary and negative,
while the ${\rm P_2}$ amplitude is complex.

\noindent
$\bbox{\rm Pseudoscalar\, exchange\, amplitudes}$.
We calculate the ($\pi,\eta$)-exchange by using the
following effective Lagrangians 
\begin{eqnarray}
&&{\cal L}_{\phi\gamma \varphi} =\frac{e}{M_\phi} g_{\phi\gamma \varphi}
\epsilon^{\mu\nu\alpha\beta}
\partial_\mu \phi_\nu \partial_\alpha A_\beta\, \varphi,\qquad
\label{L_phigammapi}\\
&&{\cal L}_{PS} =
\frac{g_{\pi^0 NN}}{2M_N}
\bar N \gamma_\mu\gamma_5\,\tau_3\, N \,\partial^\mu\pi^0
 +
\frac{g_{\eta NN}}{2M_N} \bar N \gamma_\mu \gamma_5\, N\,\partial^\mu\eta,
\label{LPS}
\end{eqnarray}
where  $\varphi=\pi^0,\eta$, and  $A_\beta$ is the photon field.
For the $\pi NN$ coupling, we use the standard value
${g_{\pi NN}}=({4\pi\cdot 14})^{1/2}$.
The status of the $\eta NN$ coupling is not so clear. The value of
$g^2_{\eta NN}/4\pi$ reported in the literatures 
varies from 0 to about 7. The largest value comes from fits to
NN scattering data, using one boson exchange potential models~\cite{Bonn}.
But many other studies~\cite{GK80,Piek93,Hats90,Tiator94}
favor  a smaller  value, $g_{\eta NN}^2/4\pi \sim 0.3- 1.0$.
In our calculation we accept the SU(3) symmetry
prediction~\cite{DeSwart}
based on the most recent value of $F/D=0.575$~\cite{F/D}. This leads to
${g_{\eta NN}}= ({0.99\cdot 4\pi})^{1/2}$
which is close to its ``overall''  averaged value.
Later, we  will discuss how our prediction of polarization observables 
can be used to pin down the value of ${g_{\eta NN}}$ .

The magnitude of the coupling constant $g_{\phi\gamma\varphi}$ in
Eq.(8) can be estimated through the decay widths~\cite{PDG} 
of $\phi \to \gamma\pi$ and $\phi \to \gamma \eta$.
We obtain $|g_{\phi\gamma\pi}|=0.14$ and $|g_{\phi\gamma\eta}|=0.707$.
Consistent with our choice of $\eta NN$ coupling constant discussed above, 
we also employ the 
SU(3) symmetry to determine their signs. We find that their 
signs are the same but are opposite to
the sign of $g_{\omega\gamma\pi}$,
which is found to be positive in the pion
photoproduction~\cite{Gari,SL}.
Since $g_{\eta pp}$ and
$g_{\pi^0  pp}$(here we include the sign due to the matrix element of
the isospin factor $\tau_3$) have the same sign,
we expect  a {\it constructive}
interference between $\pi$- and $\eta$-exchange amplitudes
in the $\gamma p\to\phi p$ reaction.
The situation changes in $\gamma n\to \phi n $ reaction.
In this case the isospin property determines that
$g_{\pi^0 nn}$($= - g_{\pi^0 pp})$ is of oppisite sign of
$g_{\eta nn}$($=g_{\eta pp}$) and hence the corresponding interference becomes
{\it destructive}.

The vertices in Fig.1b must be dressed  by form factors. 
We choose the standard form
$F_{\varphi NN}(t) =
 ({\Lambda_\varphi^2 -M^2_\varphi})/
({\Lambda_\varphi^2 -t})$ and
$F_{\phi\gamma\varphi}(t)
=
({{\Lambda_{\phi\gamma\varphi}}^2-M_\varphi^2})/
({{\Lambda_{\phi\gamma\varphi}}^2-t})$.
Here we assume that the range of the $\pi-$exchange is identical to that
determined by Friman and Soyeur~\cite{FS96} in their study of
$\rho$ and $\omega$ photoproduction. This leads us to choose
$\Lambda_\pi=0.7$ GeV and
$\Lambda^\pi_{\phi\gamma\pi} = 0.77$ GeV. There is no reliable source
to fix the cutoff parameters of the $\eta$-exchange amplitude. 
For definitness,
we set $\Lambda_{\eta} = 1$ GeV and $\Lambda_{\phi\gamma\eta}=0.9$ GeV
which are the values for the $\sigma-$exchange 
determined by Friman and Soyeur\cite{FS96} in their fit to 
the $\rho$ photoproduction.
These choices of cutoff parameters represent the main uncertainty of 
our predictions. Nevertheless, they are comparable to the values used in 
the literature and are sufficient for the present investigation. 
We find that our main conclusion concerning the isotopic effect is not 
changed unless very hard form factors( $\Lambda_\alpha > $ about 1.5 GeV) 
are used. The hard form factors are in general not favored in reproducing the
total cross sections at high energies.

\noindent
$\bbox{{\rm Direct}\, \phi-\,{\rm  radiation}\, {\rm amplitude.}}$
We evaluate this amplitude(Fig.1c) by using the
following interaction Lagrangians 
\begin{eqnarray}
{\cal L}_{\gamma NN} & = &
- e \left( \bar N \gamma_\mu\,\frac{1+\tau_3}{2}  N {A}^\mu
-\frac{\kappa_N}{2M_N} \bar N \sigma^{\mu\nu}\,N
\partial_\nu {A}_\mu \right),
\label{L_gammaNN}\\
{\cal L}_{\phi NN} & = &
- g_{\phi NN} \left( \bar N \gamma_\mu  N {\phi}^\mu
-\frac{\kappa_\phi}{2M_N} \bar N \sigma^{\mu\nu}\,N
\partial_\nu {\phi}_\mu \right),
\label{L_phiNN}
\end{eqnarray}
with $\kappa_{p(n)}=1.79(-1.91)$.
The resulting direct and crossed amplitudes in Fig.1c 
are real and their signs
are controlled by
the constants  $g_{\phi NN}$ and $\kappa_\phi$. 
Here we choose $g_{\phi NN}=-0.24$ and $\kappa_\phi=0.2$, which are 
obtained from an estimate
\cite{MMSV} of the OZI-violation contributions 
due to $\phi\pi\rho$ couplings and the
interaction of $\phi$ with the kaon cloud of the nucleon.
The calculation of Ref.\cite{MMSV} is also constrainted by SU(3) sysmmetry, 
consistent with our evaluation of the 
$(\pi,\eta)-$exchange amplitudes discussed above.
 
The 
$\phi NN$ vertices in Fig.1c may be dressed by form factors.
Namely, the direct and crossed terms 
must be dressed by $F_s^\phi$ and $F_u^\phi$ respectively.
A possible form is ~\cite{Will98,PJ91}
$F_a^\phi(a)={\Lambda^4_\phi}/({\Lambda^4_\phi + (a - M_N^2)^2}),\,
a=s,u.$
But this results in strong violation of gauge invariance with respect to
the photon and $\phi$ meson fields.
One can use the gauge invariant
prescription of \cite{Hab97}, where a universal form factor
$F^\phi(s,u)=\frac12 (F^\phi_s(s) +  F^\phi_u(u))$ is
used to regularize both $\phi NN$ vertices.
But our predictions would then have an additional uncertainty\cite{Hab97} 
depending on the cutoff parameter $\Lambda_\phi$.
To avoid this problem we use the prescription of \cite{Ohta} and
simply set $F^\phi_s = F^\phi_u=1$.
Our prediction of the contribution from the direct
$\phi-$radiation(Fig.1c) may therefore be considered as an upper limit.

\noindent
${\bbox {\rm Results.}}$
 With the ($\pi, \eta$)-exchange and direct $\phi$-radition amplitudes 
defined above, 
our first task is to determine the parameters of Pomeron-exchange
amplitudes. The parameters $s_i$ and $C_i$ of Eq.(4) are not independent in
the fit to the data. For example, a smaller value of $s_2$ will
lead to a larger $C_2$ in the fit. For simplicity we set $s_2=0$.
To account for the threshold behavior of the P$_1$  amplitude,
we set $s_1= \sqrt{M_N+M_\phi}$.
We then find that the $d\sigma/dt(\theta=0)$
data of $\gamma p \rightarrow \phi p$ reaction can be reproduced with
$\xi = C_2/C_1 = -0.55$ and $C_1=2.34$. Our result is the thick solid curve
in the left panel of Fig.3. The slope $B=1.7$ GeV$^{-2}$
of the form factor Eq.(6)
is needed to reproduce the angular distribution,
as shown in the right panel of Fig.3. Here we also display the
relative importance between different amplitudes. We note that the
P$_1$ contribution drops sharply as energy approaches the production
threshold. On the other hand, the P$_2$ contribution increases
very quickly as energy decreases. This striking difference
in energy-dependence between P$_1$ and P$_2$ is due to the different signs of
their intercepts: $\alpha_1(0)=1.08$ and $\alpha_2(0)=-0.75$(see Eq.(5)).
By comparing the $\pi$-curve and $(\pi,\eta)$-curve in Fig.3, we see 
that the $\pi$ and $\eta$ amplitudes
interfere constructively. This is exactly what we expected in the discussion
given above. The direct $\phi$-radiation clearly dominates the cross
sections at large angles.

The fits to the $\gamma p \to \phi p$ data completely fix the model and we
can now apply the model to predict the $\phi$ photoproduction on neutron.
Our results for $\gamma n \rightarrow \phi n$
are shown in Fig.4. For comparison, we also display the corresponding
$\gamma p \rightarrow \phi p$ results. Here
we observe that $\pi-$ and $\eta-$exchange interfere destructively(comparing
the $\pi$-curve and $(\pi,\eta)$-curve). The direct $\phi$-radiation
contribution(dash-dotted curves) is drastically different from that in
Fig.3. This is due to the  large difference between the $\gamma pp$ and
$\gamma nn$ vertices. 
However, the change in the cross sections due to this
isotopic effect is very small except at
large scattering angles (comparing the top two curves in Fig.4) where the
cross sections are very small. Experimental verification of this prediction
is very difficult.  

We now point out that the isotopic effect discussed above can be verified by
examining polarization observables. As an example, we show in
Fig.5 our prediction of the $\phi$ decay asymmetry~\cite{SSW}
\begin{eqnarray}
\Sigma_{\phi} = \frac{\sigma_{\parallel}-\sigma_{\perp}}
{\sigma_{\parallel}+\sigma_{\perp}} ,
\end{eqnarray}
where $\sigma_{\parallel}$($\sigma_{\perp}$) are
the cross sections induced by photons polarized in the direction
parallel (perpendicular) to the $\gamma-\phi$ plane.
These two cross sections refer to measurements of
the symmetric $K^+K^-$ pairs produced in the $\phi$ decay.
Our prediction of $\Sigma_{\phi}$ at $E_\gamma = 2$ GeV are shown
in Fig.5. We find that the predicted isotopic effect at forward
angles is largely due to the $\pi -\eta$ interference.
This can be understood by comparing the ($\pi,\eta)-$curve and $\phi NN$-curve
in the right panels of Figs. 3 and 4.
Consequently, the predicted energy-dependence at $\theta=0$ 
shown in the left panel of Fig.6 can be used to explore
the relatively undetermined $\eta NN$ coupling constant.
To facilitate such a study, 
we show the dependence of $\Sigma_\phi (\theta=0)$ on
$g_{\eta NN}$ in the right panel of Fig.6.
In this calculation,
the strength $C_2$ of the $P_2$ Pomeron is adjusted to maintain the same fit to
the $\gamma p \rightarrow \phi p$ data as $g_{\eta NN}$ is varied in
the calculation. 
At large angles, the cross sections are dominanted by the direct
$\phi-$radiation amplitude. Hence a measurement of $\Sigma_{\phi}$ at
large angles can be used to study the $\phi NN$ coupling.
However, such a measurement may be very difficult because the cross sections
become very small at large angles.

In summary, we have developed a model to describe the $\phi$ photoproduction
at low energies, $\sqrt{s}\sim 2 - 5$ GeV. In addition to the traditional
Pomeron trajectory  with an intercept of $\alpha_1(0)=1.08$, the model
contains a second Pomeron trajectory with a negative intercept
$\alpha_2(0) = -0.75$ inspired by the ($J^{\pi} =0^+$,
$M^2_{\rm gb} \sim 3$ GeV) glueball  predicted by Lattice QCD calculations
and a QCD-based Dual Landau-Ginsburg model~\cite{LGM}. The model also
includes the contributions from
($\pi$, $\eta$)-exchange and direct $\phi$-radiation from the nucleon.
We have found a strong isotopic effect in the predicted
$\phi$ decay asymmetry (Figs.5, 6), indicating the possibility of
using the $\phi$ photoproduction polarization observables
to explore the $\eta NN$ and $\phi NN$ couplings. Experiments
for measuring such observables at LEPS of
SPring-8 (Japan) and TJNAF (USA) are highly desirable.


$\bbox{\rm Acknowledgments}.$
We gratefully acknowledge  useful discussions with
M.~Fujiwara,
T. Hotta,
T.~Morii,
T.~Nakano,
Y.~Oh,
E.~Oset and
D.~Roy.
A.I.T. appreciates support of the COE professorship program and
the warm hospitality of the Research Center for Nuclear
Physics  of  Osaka University. This work is also partially supported
by the U.S. Department of Energy, Nuclear Physics Division, under
Contract NO. W-31-109-ENG-38.

%
%
%
%
%
%
%
%
%

\begin{figure}[h]
\centering
\epsfig{file=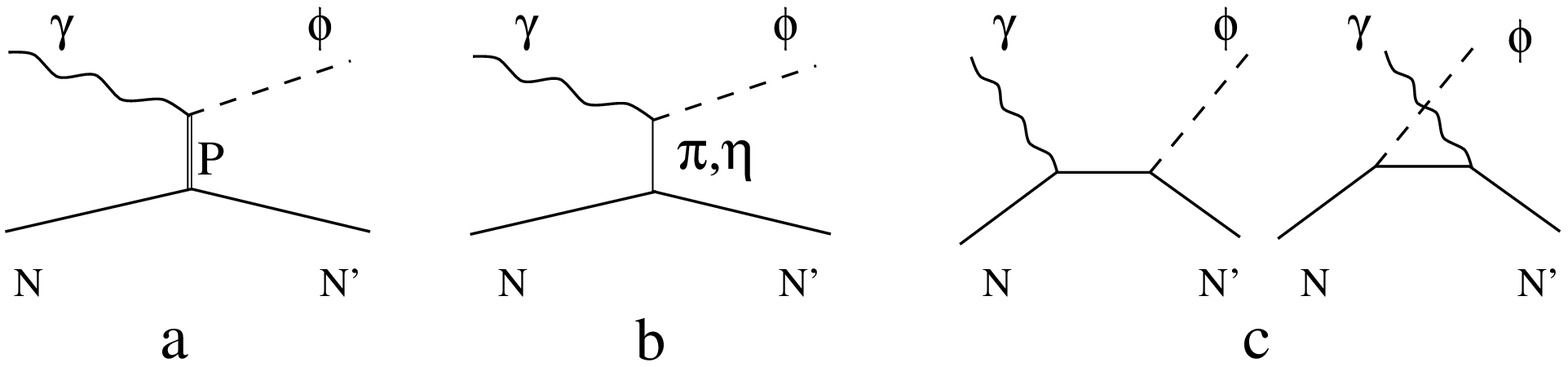, width=12cm}
\vspace*{0.5cm}
\caption{Diagrammatic representation of 
$\phi$ production mechanisms:  (a) Pomeron-exchange,
(b) ($\pi,\eta$)-exchange, and (c) direct $\phi-$radiation.}
\label{Fig1}
\end{figure}

\begin{figure}[h]
\centering
\epsfig{file=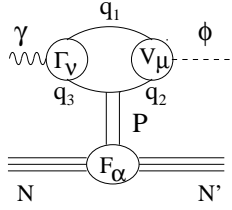, width=3cm}
\vspace*{0.5cm}
\caption{Diagrammatic representation of the Donnachie-Landshoff 
model of Pomeron exchange}
\label{Fig2}
\end{figure}

\begin{figure}[h]
\centering
\epsfig{file=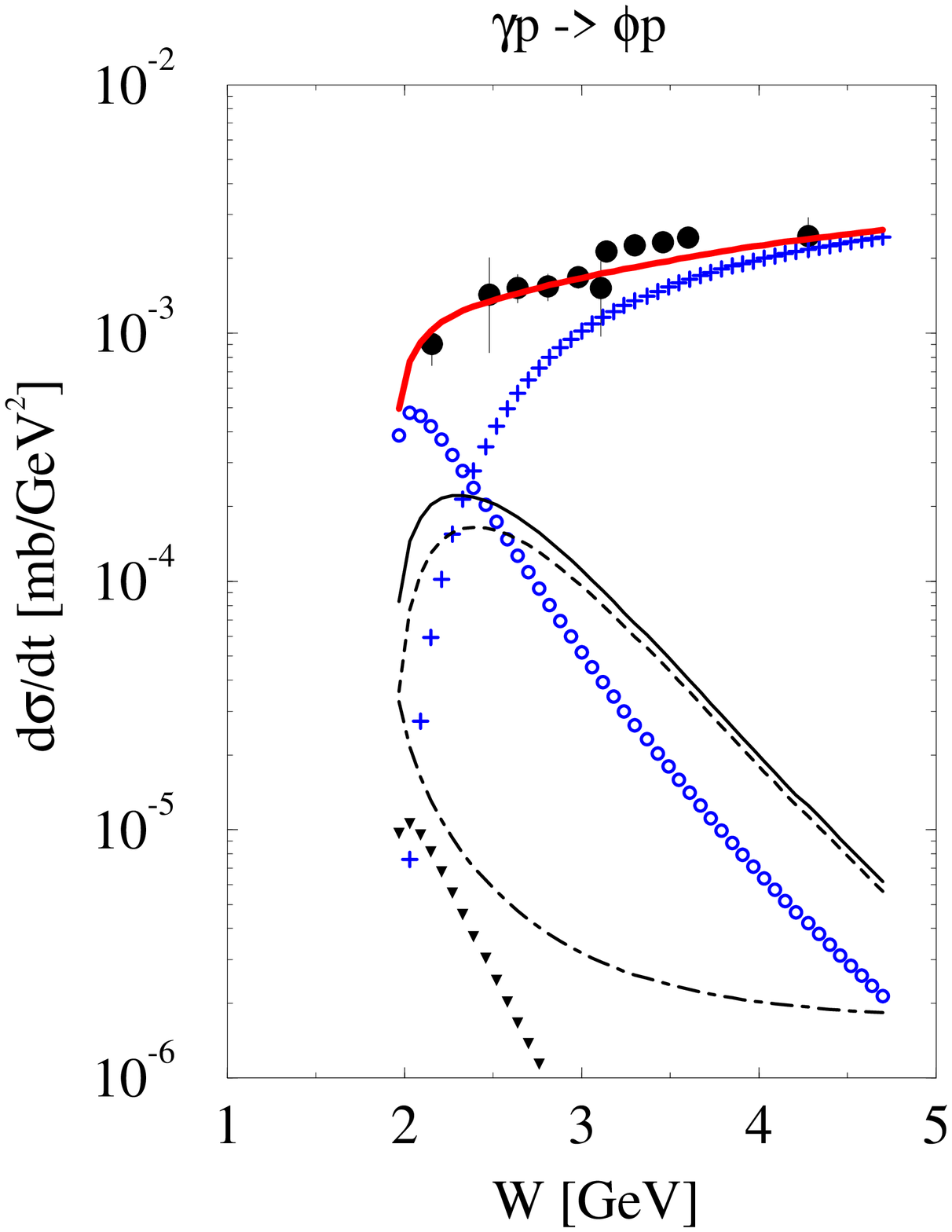, width=7cm}\qquad\qquad
\epsfig{file=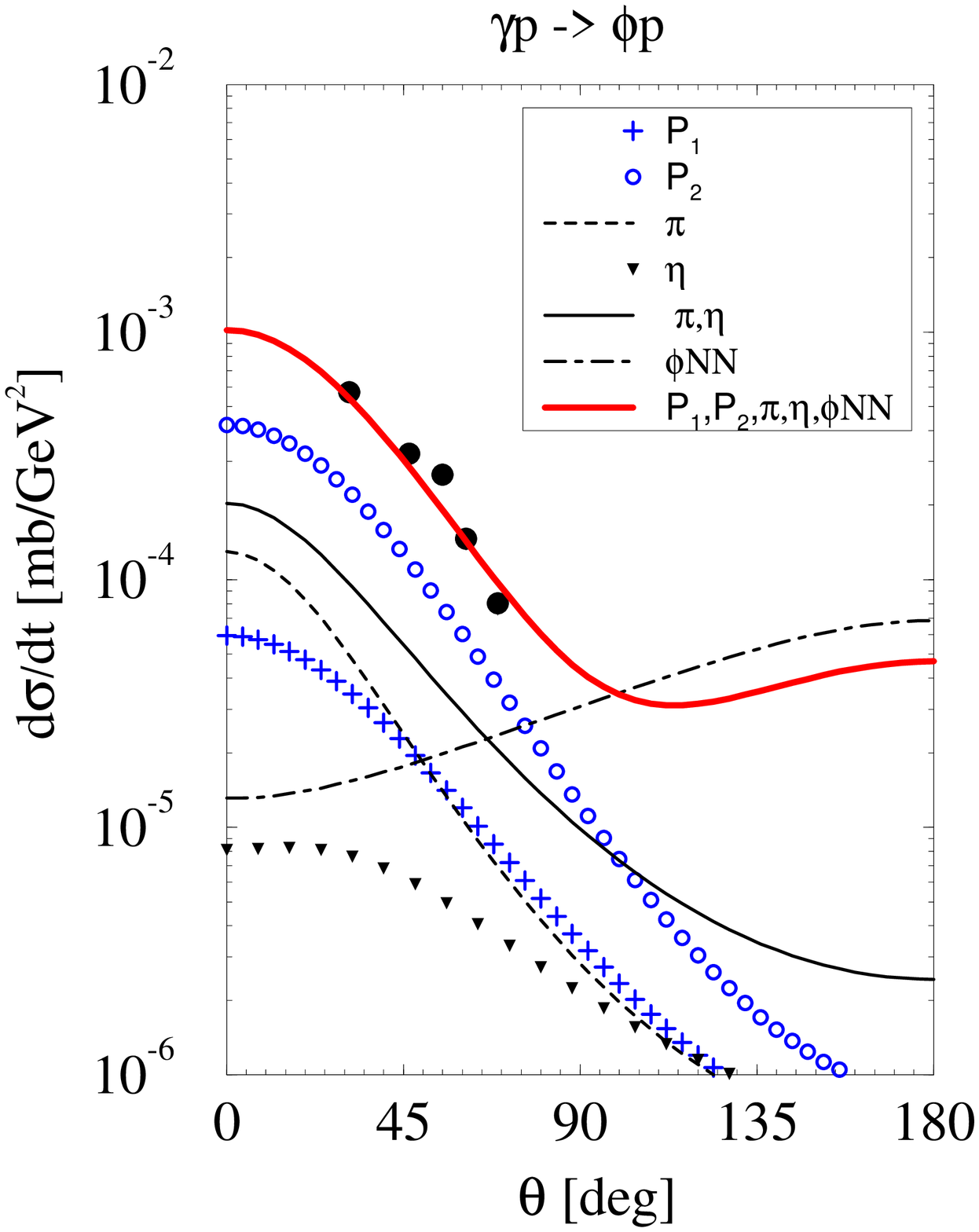, width=7cm}
\vspace*{0.5cm}
\caption
{Right panel:differential cross section for the 
$\gamma p\to p\phi$ reaction at $t=t_{\rm max}$ ($\theta=0$)
 as a function of the total energy
energy distribution $W=\sqrt{s}$.
Data are taken from \protect\cite{DATA1,DATA2,DATA3}
Right panel: differential cross section 
at $E_\gamma=2$ GeV ($W$=2.15 GeV).
Data are taken from~\protect\cite{DATA1}.}
\label{Fig3}
\end{figure}

\begin{figure}[h]
\centering
\epsfig{file=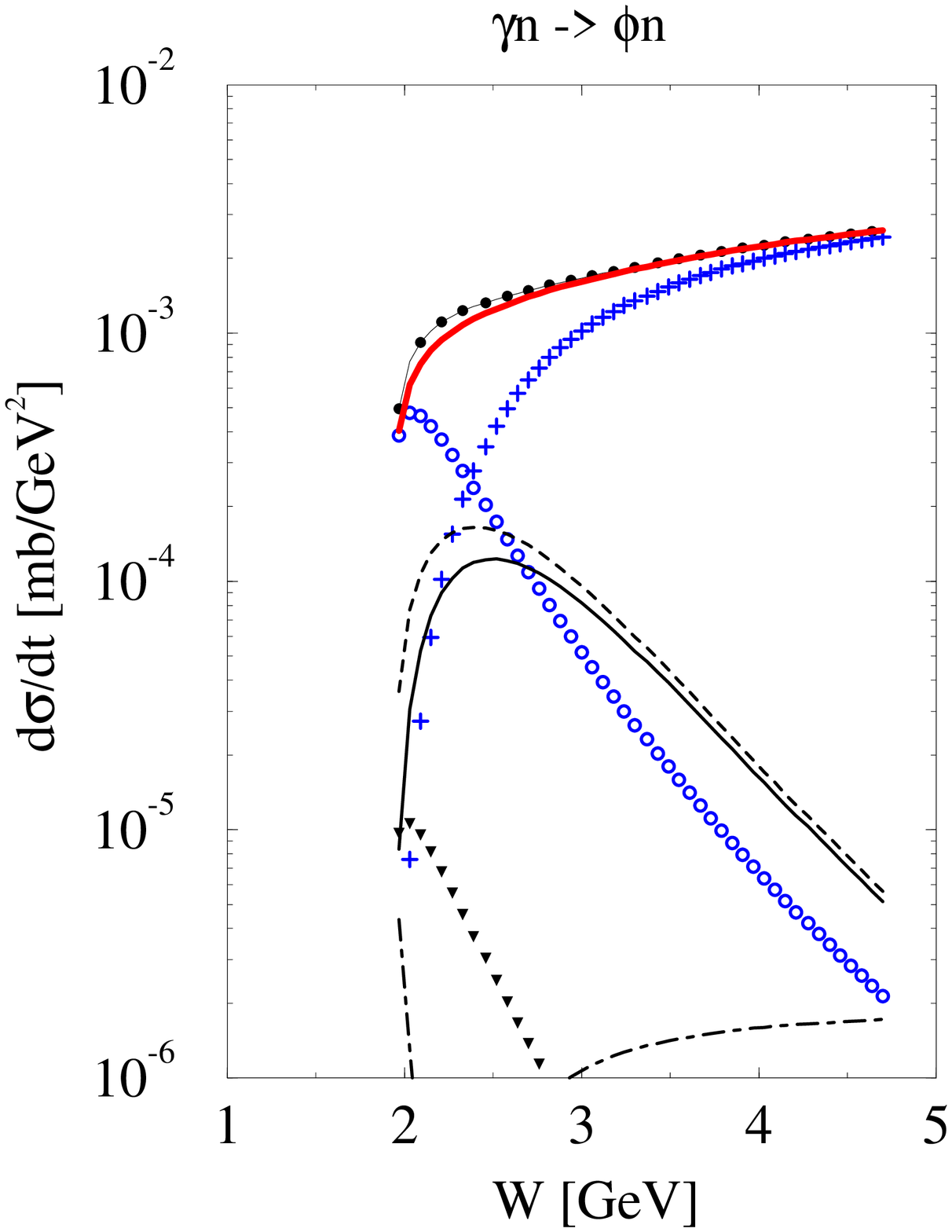, width=7cm}\qquad\qquad
\epsfig{file=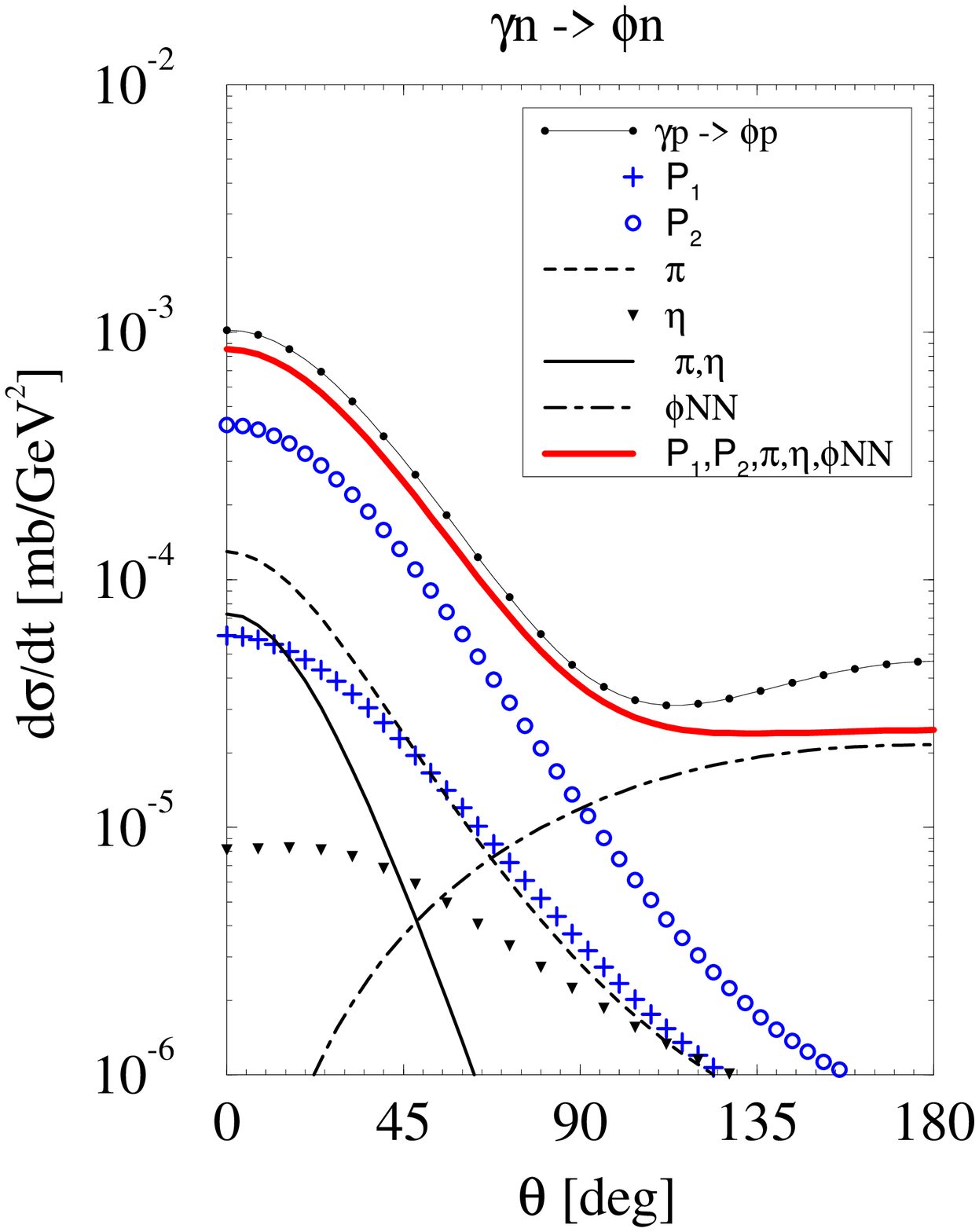, width=7cm}
\vspace*{0.5cm}
\caption
{Same as Fig.3 except for the $\gamma n\to n\phi$ reaction.}
\label{Fig4}
\end{figure}

\begin{figure}[h]
\centering
\epsfig{file=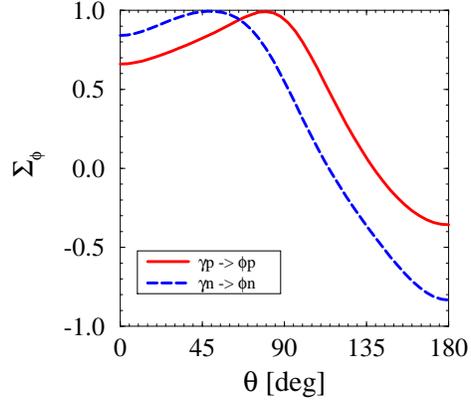, width=6cm}
\vspace*{0.5cm}
\caption{The $\phi$ decay asymmetry $\Sigma_\phi$, defined in Eq.(13),
as function of $\theta$ at $E_\gamma$=2 GeV.}
\label{Fig5}
\end{figure}

\begin{figure}[h]
\centering
\epsfig{file=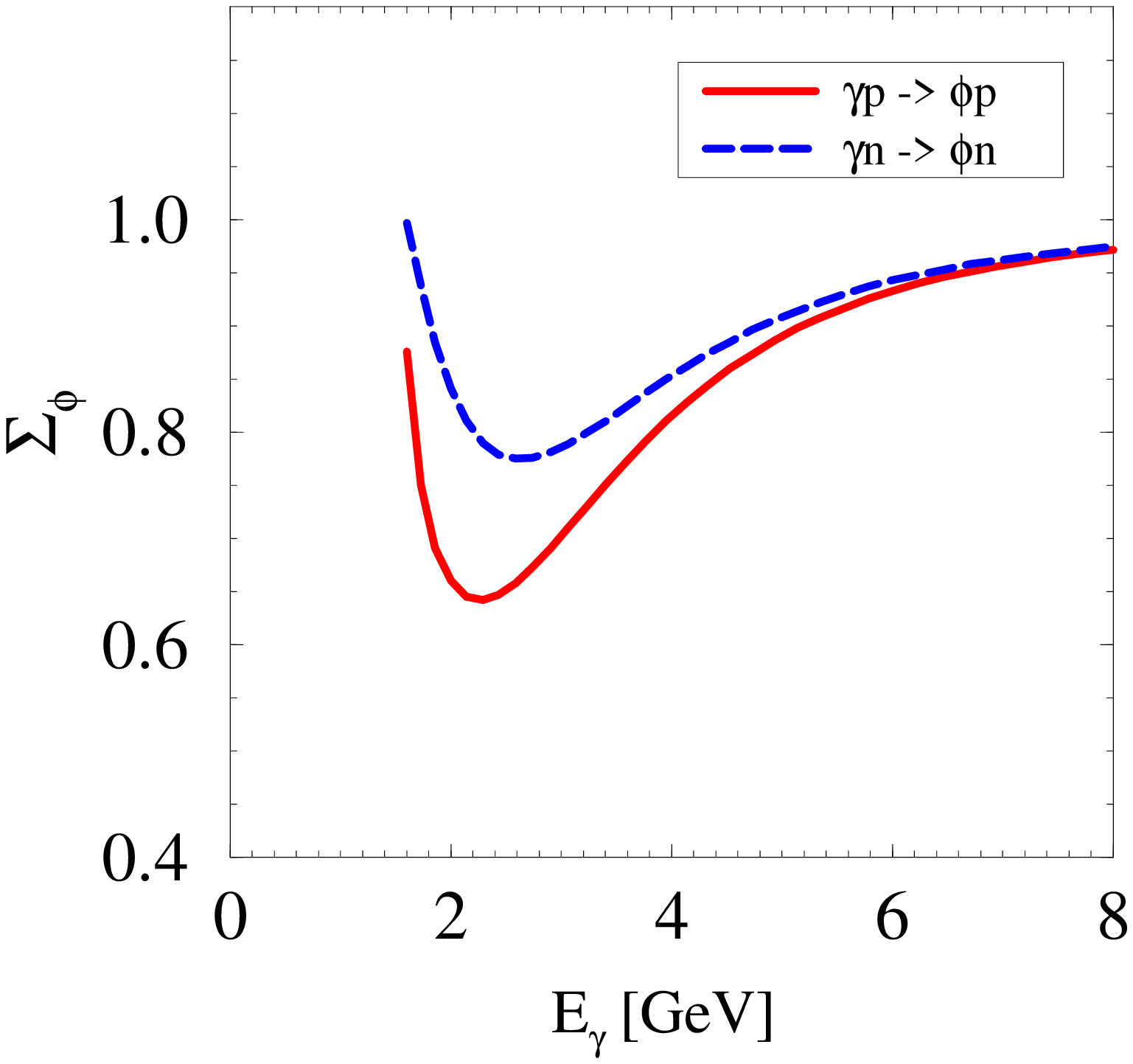, width=6cm}\qquad\qquad
\epsfig{file=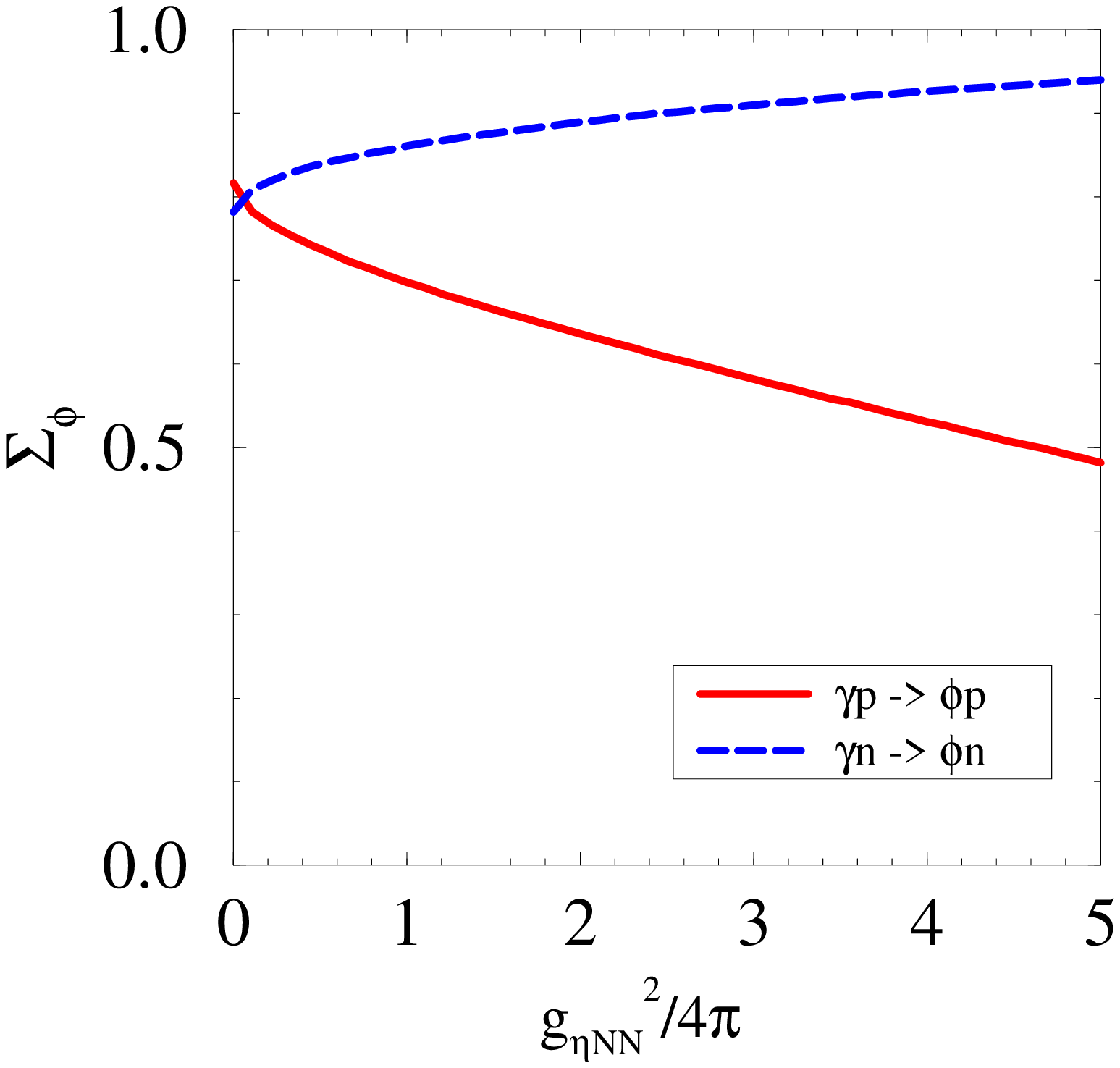, width=6cm}
\vspace*{0.5cm}
\caption{The $\phi$ decay asymmetry $\Sigma_\phi$, defined in Eq.(13), 
at $\theta=0$ as a function of the
photon energy (left panel) and as
a function of  the $\eta NN$ coupling constant (right panel).}
\label{fig:6}
\end{figure}

\end{document}